\documentclass[prl,twocolumn,preprintnumbers]{revtex4}

\usepackage{times}

\usepackage{bm}
\usepackage{graphicx}
\usepackage{amsbsy}
\usepackage{amsmath}
\usepackage{amsfonts}
\usepackage{amsthm}

\begin{document}

\theoremstyle{plain}
\newtheorem{theorem}{Theorem}
\newtheorem{lemma}[theorem]{Lemma}
\newtheorem{corollary}[theorem]{Corollary}
\newtheorem{proposition}[theorem]{Proposition}
\newtheorem{conjecture}[theorem]{Conjecture}

\theoremstyle{definition}
\newtheorem{definition}[theorem]{Definition}

\title{Purification of Mixed State with Closed Timelike Curve is not Possible}
\author {Arun K. Pati, Indranil Chakrabarty and Pankaj Agrawal}
\affiliation{Institute of Physics, Sainik School Post,
Bhubaneswar-751005, Orissa, India}

\date{\today}

%\maketitle

\begin{abstract}
In ordinary quantum theory any mixed state can be purified in an enlarged
Hilbert space by bringing an ancillary system. The purified state does not 
depend on the state of any extraneous system with which the mixed state 
is going to 
interact and on the physical interaction.
Here, we prove that it is not possible to purify a mixed state that traverses 
a closed time like curve (CTC) and allowed 
to interact in a consistent way with a causality-respecting (CR) quantum system
in the same manner.
%In other words, if a CTC quantum system with mixed state has to 
%undergo arbitrary interaction with arbitrary CR system then 
%it is always in a `proper mixture', as 
%it cannot be regarded as
%subsystem of pure entangled state. 
Thus, in general for arbitrary 
interactions between CR and CTC systems 
there is no universal `Church of the larger Hilbert space' for mixed states 
with CTC. This shows that in  
quantum theory with CTCs there can exist `proper' and `improper' mixtures.
 \end{abstract}

\maketitle

For many years physicists believed that the existence of closed
timelike curve (CTC) \cite{mor,fro,kim,got,haw} is only a theoretical
possibility rather than a feasibility. A closed time like curve
typically connects back on itself, for example, in the presence
of a spacetime wormhole that could link a future spacetime point
with a past spacetime point. However, there have been 
criticisms to the existence of CTCs and the ``grandfather paradox'' 
is one. But Deutsch proposed a computational model of quantum systems in 
the presence of CTCs and 
resolved this paradox by presenting a method for finding
self-consistent solutions of CTC interactions in quantum theory
\cite{deu}. He has also investigated various quantum
mechanical effects that can occur on and near the closed timelike curves,
including the possibility of violation of the correspondence
principle and unitarity. This has opened up a new area of
research in recent times. In quantum information theory some authors 
have assumed
that CTCs exist and examined the consequences of this assumption
for quantum computation \cite{bru,bac,aar}. Several authors have
come up with the idea that access to CTCs could enhance the
computational power. For example, in reference \cite{bru} it was
proposed that access to a CTC would factor composite numbers
efficiently with the help of a classical computer only. This
clearly gave strong indications that the CTC-assisted
computers may be more powerful. The author in reference
\cite{bac} has moved one step forward and argued that a CTC-assisted
quantum computer would be able to solve NP-complete problems,
which is an impossibility for a quantum computer alone. Another 
result about computational complexity is that the power of a
polynomial time bounded computer in classical or quantum world 
assisted by CTCs can be solved in a space polynomial in the problem size 
(PSPACE) but potentially exponential time \cite{aar}.

In an interesting work Brun {\em et al} \cite{bru1} have shown how a party
with access to CTCs can perfectly
distinguish nonorthogonal quantum states. The result came as a
shock to the community as it put fundamental quantum
cryptographic protocols (like BB84 \cite{ben} protocol) at stake.
This is so because the encrypting scheme that depends on the
principles of quantum theory can be broken by allowing the
quantum message to interact with another quantum system that
travel through a CTC region. At the same time this raises several 
questions about the nature of the quantum world with 
closed timelike curves. 
%Because, their existence challenges 
%the basic principle of quantum mechanics that nonorthogonal states cannot be
%perfectly distinguished.

However, Bennett {\it et al} have a completely opposite view
to this proposal \cite{ben1}. They studied the power of closed
timelike curves (CTCs) for the tasks of distinguishing
nonorthogonal input states and speeding up otherwise hard
computations. They showed that when a CTC-assisted computer is
feed up with a labelled mixture of states as an input it can do no better
than it would have done without CTC assistance. The conclusions
drawn in \cite{bru1} are based on considering only fixed input
states which are pure. These conclusions do not hold for general
case where inputs consist of mixed states. They resolved the
contradiction between their claim and previous results by 
arguing that as the CTC-assisted evolution is not linear, the
output of such computation on a mixture of inputs is not a convex
combination of its output on the pure components of the mixture.
Recently, \cite{ralph}, it has been argued that by keeping 
track of the information flow of quantum system during the interaction with CTCs
one can have a better picture of the Deutsch model and this also supports
the notion that CTC assisted party can distinguish non-orthogonal quantum 
states as reported in \cite{bru1}.
This once again suggests that the whole situation is worth rethinking 
about the true power of CTC world as there is
no direct experimental evidence whether having access to the CTC qubit or
nonlinearity do provide any help in this regard. It may be worth mentioning 
another strange feature 
of the Deutsch model, namely, that the dynamical evolution 
through the region containing CTCs become discontinuous function of the 
input state \cite{imbo}.

Now coming back to Deutsch formalism \cite{deu}, it involves a
unitary interaction between a causality-respecting (CR) quantum system
with another system that is having a world line in the region 
of a closed timelike curve (CTC). It is assumed that these states 
are quite similar to the density matrices that we have in standard quantum
mechanics. Before the interaction the combined state of the CR system 
and CTC system is in a product state $\rho_{\rm CR}\otimes \rho_{\rm CTC}$.
The unitary interaction between the CR and CTC systems 
transforms the initial composite state as
\begin{equation}
\rho_{\rm CR}\otimes \rho_{\rm CTC} \rightarrow 
U (\rho_{\rm CR}\otimes \rho_{\rm CTC}) U^{\dagger}. 
\end{equation}
For each initial mixed 
state $\rho_{\rm CR}$ of the CR
system, there do exists a CTC system $\rho_{\rm CTC}$ such that after the 
interaction we must have 
\begin{equation}
{\rm  Tr}_{\rm CR}(U \rho_{\rm CR}\otimes \rho_{\rm CTC} U^{\dagger})=\rho_{\rm CTC}.
\end{equation}
This is the self-consistency condition in the Deutsch model.
Since the time travelled system follows a CTC, travelling back in time 
will make the input density matrix equal to its output density matrix.
Mathematically, the solution to this equation is a fixed point
(there may be multiple fixed points, in which case one has to take the
maximum-entropy mixture of them) \cite{deu}. The final state of the 
CR quantum system is then defined as
\begin{equation}
{\rm Tr}_{\rm CTC}(U \rho_{\rm CR}\otimes \rho_{\rm CTC} U^{\dagger}) =
\rho_{\rm CR}^{'}.
\end{equation}
In a nutshell these are the two basic conditions which actually govern
the entire unitary dynamics of the CTC and CR quantum systems.
Here, one should note that $\rho_{\rm CTC}$ depends nonlinearly 
on $\rho_{\rm CR}$ and hence the output of the CR system is a nonlinear 
function of the input density matrix \cite{cas}.
% because $\rho_{\rm CTC}$ depends on the initial state $\rho_{\rm CR}$.

In this paper we ask what should be the nature of the density 
operator of a quantum system that traverses a closed time like curve. 
Is it similar in all respect to causality-respecting quantum system? 
In particular, we ask can we always purify $\rho_{\rm CTC}$ that interacts 
with a CR system indepedent of anything else? Quit surprisingly, we 
show that there does not exist 
general purification for a mixed state density operator of a 
CTC quantum system in the Deutsch model for arbitrary $\rho_{\rm CR}$ and $U$. 
In other words, given a CTC system in the Deutsch formalism which is 
in a mixed state, if it has to interact with arbitrary CR system and subject to
arbitrary interaction, then that cannot be viewed as a
subsystem of a universal pure entangled state in an enlarged Hilbert space of
the composite CTC system (up to local unitary freedom). 
%Therefore, any mixture that traverses a CTC and interacts with a CR system 
%has to be a `proper mixture'. This also suggests that `improper mixtures' with 
%CTC world line cannot interact with CR systems. 
Since, in quantum information theory purification of mixed states finds many
important applications, we believe that our result  
have deep implications for quantum theory in the presence of CTCs.
Recently, it was shown that in general probabilistic theories that allow
for purification one can obtain most of the quantum formalism and 
principles of quantum information processing \cite{chiri}. This may 
provide a strong motivation to our result, 
showing that closed timelike curves exclude the universal purification 
implies that they exclude a central piece of quantum mechanics.

In quantum theory the density operator $\rho \in {\cal B}({\cal H})$, where 
${\cal B}({\cal H})$ is the set of density operators on a given Hilbert space 
${\cal H}$. It is a Hermitian, positive and trace class operator 
with ${\rm Tr}(\rho) =1$. 
In the ordinary quantum world there is no distinction between
a `proper mixture' and an `improper mixture'. The {\it proper mixture} $\rho$ is
a convex combination of subensembles of pure states $|\psi_k\rangle$, each
occurring with a probability $p_k$. Thus, $\rho$ can be expressed as 
$\rho = \sum_k p_k |\psi_k\rangle
\langle \psi_k |$  and this bears the ignorance interpretation. 
The above decomposition is not unique and can be expressed in 
infinitely many ways as a convex sum of distinct, but not necessarily 
orthogonal projectors. 
This notion was introduced by Von Neumann when we do not
know the exact preparation procedure in describing the physical system.  
%Given a particular decomposition for $\rho$, if $|\psi_k\rangle$'s are 
%pairwise orthogonal then the above will be a spectral decomposition. 
%This is unique if and only if there is no degeneracy. When there is 
%degeneracy or if the pure states are not eigenvectors of $\rho$, 
%then we have infinite number of decompositions.
The {\it improper  mixture} corresponds to the result of tracing out of
a pure projector of a composite system (system plus ancilla) 
$S+A$ such that $\rho_S = {\rm Tr}_A
(|\Psi \rangle_{\rm SA}\langle \Psi |)$, where 
$|\Psi \rangle_{\rm SA} = \sum_k \sqrt{p_k}|\psi_k\rangle_S |\phi_k\rangle_A$. 
However, there is no way to differentiate
a proper mixture from improper one. This is so, because, given any density
matrix we can always purify it in an enlarged Hilbert space (in infinite
number of ways). In fact, it has been argued that in the usual quantum theory
all the density matrices can be regarded as `improper mixture' \cite{kirk}.
Most importantly, the pure entangled state does not depend on the state of 
other extraneous system with which $\rho_{\rm S}$ is going to interact nor it 
depends on the interaction that might take place between the system and anything
else. However, this is not the case for CTC quantum systems. 

%Below we prove the no-purification theorem for CTC system that is allowed 
%to interact with CR system in the Duetsch model.

{\bf Theorem I:}
Let $\rho_{\rm CTC} =
\sum_k p_k |k\rangle\langle k| $ be the spectral decomposition
of the CTC mixed state that interacts with a CR quantum system in a pure state. 
For arbitrary $U$ on ${\cal H}_{\rm CR} \otimes {\cal H}_{\rm CTC}$
and $|\psi\rangle_{\rm CR} \langle \psi|$ on ${\cal H}_{\rm CR}$ if the 
CTC density matrix satsifies the consistency condition 
$\rho_{\rm CTC} = {\rm  Tr}_{\rm CR}(U \rho_{\rm CR}\otimes \rho_{\rm CTC} 
U^{\dagger}) $ then there does not exist general purification for 
$\rho_{\rm CTC}$, such that $\rho_{\rm CTC} = {\rm Tr}_{\rm CTC'}
[|\Phi \rangle \langle \Phi |]$, where
$|\Phi\rangle_{\rm CTC, CTC'}=\sum_k \sqrt{ p_k}|k\rangle_{\rm CTC}
|b_k\rangle_{\rm CTC'}$ for all $\rho_{\rm CR}$ and $U$.
%with $|\Phi \rangle$ being indepedent of $\rho_{\rm CR}$ and $U$.
\\

{\bf Proof:}
We start with two quantum systems as inputs to the 
unitary evolution, where one system is from the causality respecting (CR)
region and another from the causality violating region. 
Let these density matrices are given by 
$\rho_{\rm CR}=|\psi\rangle_{\rm CR} \langle \psi|$ and 
$\rho_{\rm CTC}=\sum_k p_k|k \rangle_{\rm CTC} \langle k|$.
Now, they undergo unitary evolution in the Deutsch model according to (1).
%\begin{align}
%\rho_{\rm CR}\otimes \rho_{\rm CTC} \rightarrow 
%U (\rho_{\rm CR}\otimes \rho_{\rm CTC}) U^{\dagger}.
%\end{align}
Let us assume that the purification is possible for the CTC system. 
Then, it can be thought of as a part of a pure entangled 
state of a composite system in the Hilbert space 
${\cal H}_{\rm CTC} \otimes {\cal H}_{\rm CTC'}$.
This pure entangled state  can be written in
the Schmidt decomposition form as (up to local unitary in the Hilbert space of 
${\rm CTC'}$)
\begin{eqnarray}
 |\Phi\rangle_{\rm CTC,CTC'} = 
\sum_k\sqrt{ p_k}|k\rangle_{\rm CTC}|b_k\rangle_{\rm CTC'}
\end{eqnarray}
with $\rho_{\rm CTC} = {\rm Tr}_{\rm CTC'} [|\Phi \rangle \langle \Phi |]$.
Here, $p_k$'s are the Schmidt coefficients, $|k \rangle$'s are orthonormal 
basis for the CTC Hilbert space and $|b_k\rangle$ are orthonormal 
basis for the ${\rm CTC'}$ Hilbert space.
In the pure state picture the unitary evolution is equivalent to 
\begin{align}
%\begin{eqnarray}
& |\psi \rangle_{\rm CR}\langle \psi| \otimes |\Phi\rangle_{\rm CTC,CTC'} 
 \langle \Phi|  \rightarrow  \nonumber\\
& U \otimes I (|\psi \rangle_{\rm CR} \langle \psi| \otimes 
|\Phi\rangle_{\rm CTC,CTC'} 
 \langle \Phi| ) U^{\dagger} \otimes I,
%\end{eqnarray}
\end{align}
where $U$ acts on the Hilbert space of CR and CTC and $I$ acts on 
${\rm CTC'}$. If the purification holds then after the evolution the 
combined state of the CR and CTC system is given by
\begin{align}
{\rm Tr}_{\rm CTC'} [U \otimes I (|\psi \rangle_{\rm CR} \langle \psi| \otimes 
|\Phi\rangle_{\rm CTC,CTC'} 
 \langle \Phi| ) U^{\dagger} \otimes I ] = \nonumber\\
= \sum_k p_k U(|\psi \rangle_{\rm CR} \langle \psi| \otimes 
|k \rangle_{\rm CTC} \langle k |)U^{\dagger}.
\end{align}
From (1) and (6) we have 
\begin{align}
 U(|\psi \rangle_{\rm CR} \langle \psi| \otimes 
\sum_k p_k |k \rangle_{\rm CTC} \langle k |)U^{\dagger}= \nonumber\\ 
\sum_k p_k U(|\psi \rangle_{\rm CR} \langle \psi| \otimes 
|k \rangle_{\rm CTC} \langle k |))U^{\dagger}.
\end{align}
Now taking the partial trace over the CR system and using the 
consistency condition we have 
\begin{align}
\sum_k p_k |k \rangle_{\rm CTC} \langle k | = 
{\rm Tr}_{\rm CR} [\sum_k p_k U(|\psi \rangle_{\rm CR} \langle \psi| \otimes 
|k \rangle_{\rm CTC} \langle k |))U^{\dagger} ].
\end{align}
Now the action of the unitary operator on the CR system and  
the orthonormal basis of the CTC system can be written in the Schmidt 
decomposition form as
\begin{align}
 U(|\psi\rangle_{\rm CR}|k\rangle_{\rm CTC} )= \sum_{m}\sqrt{f_m^k(\psi)}
|a_m^k(\psi)\rangle_{\rm CR}  |u_m(k) \rangle_{\rm CTC},
\end{align}
where $f_m^k(\psi)$'s are the Schmidt coefficients with 
$\sum_m f_m^k(\psi) =1$ for all $k$, $|a_m^k(\psi)\rangle_{\rm CR}$ and 
$|u_m(k) \rangle_{\rm CTC}$ 
are orthonormal Schmidt bases for the CR system and CTC system, respectively.
%It may be noted that for any non-trivial interaction between the CR system 
%and the CTC system, we must have non-zero values of $f_m(k)$ at least for two 
%cases.
Therefore, from (8) we have 
\begin{eqnarray}
\sum_k p_k|k \rangle\langle k| =\sum_{km} p_k f_m^k(\psi) |u_m(k) \rangle \langle
 u_m(k)|.
\end{eqnarray}
From (10), the eigenvalues of $\rho_{\rm CTC}$ are given by
\begin{eqnarray}
p_n = \sum_{km} p_k f_m^k(\psi) |\langle n |u_m(k) \rangle |^2
\end{eqnarray}
which depends on $\rho_{\rm CR}$ and $U$. This is a recursive equation. 
However, the spectrum of $\rho_{\rm CTC'}$ is independent of $\rho_{\rm CR}$ 
and $U$ (as it does not interact with anything and in principle can be far away 
from the region where CR and CTC systems interact). Since we must have the 
equal spectrum for any pure bipartite entangled state, 
i.e., Spec($\rho_{\rm CTC}$) $=$ Spec($\rho_{\rm CTC'}$), the purification 
assumption and consistency condition violate this.
Thus, for any non-trivial interaction between the CR and the CTC to happen, 
our assumption that the purification is
possible for $\rho_{\rm CTC}$ must be wrong. Hence, there is  no 
purification of a CTC mixed state for arbitrary CR system and arbitrary 
interaction.

Our proof also rules out the possibility that a CTC mixed state can be 
purified by attaching a CR system as part of the purification. Even if we 
imagine that CTC can be
purified with a CR system, then similar arguments will go through 
and we will get a contradiction. Thus, {\em there is no universal 
`Church of the larger Hilbert space' (either in CR or CTC world) where a 
CTC mixed state can be
purified} if it has to undergo arbitrary interaction with arbitrary CR 
system in a non-trivial way. It will be interesting to see if the 
no-purification 
theorem holds in other non-linear quantum theories \cite{akp}.

%Therefore, if a CTC system with a density matrix $\rho_{\rm CTC}$ 
%has to interact with a CR system then that must be a `proper mixture'.
%This also proves that there can be no other degrees of freedom for CTC 
%system which when ignored will give the density matrix $\rho_{\rm CTC}$ 
%and can interact with a CR system.

Does our result rule out the existence of `improper' mixture?
The answer is no. If already we have a composite CTC system in a 
pure entangled state, then the subsystems will be in a mixed state and 
that will be a `improper mixture'.
%However, a CTC system with an `improper mixture' if it interacts with arbitrary
%CR system then it cannot do so in a consistent manner within 
%the Deutsch model. 
As long as the part of the entangled state 
of a CTC system does not interact with a CR system in a pure state, 
the existence of `improper mixture' is allowed. One may aruge that 
for a fixed known 
$\rho_{\rm CR}$ and $U$ one can purify $\rho_{\rm CTC}$. But then the 
pure entangled state depends on $\rho_{\rm CR}$ and $U$, i.e., $|\Phi \rangle = 
|\Phi(\psi, U) \rangle $.
This is in sharp distinction to the purification in ordinary quantum theory:
If we have two systems (say) with density matrices $\rho$ and $\rho_{\rm S}$ 
and they interact via $\rho \otimes \rho_{\rm s} \rightarrow 
U (\rho \otimes \rho_{\rm s}) U^{\dagger}$, then we can always purifiy 
$\rho_{\rm S}$ such that $\rho_S = {\rm Tr}_A
(|\Psi \rangle_{\rm SA}\langle \Psi |)$, where 
$|\Psi \rangle_{\rm SA}$ is the purified state and it 
does not depend on $\rho$ and $U$. The purification of 
a density matrix in ordinary quantum theory is universal.
However, in the case of CTC quantum theory it is not so.
Also, in the case of CTC quantum theory, 
it is not clear why the eigenvalues of $\rho_{\rm CTC'}$ which may be far 
away depends on 
the $\rho_{\rm CR}$ and $U$ which it has not seen. There seems to be a 
problem with the locality principle in quantum theory with CTCs. 
Nevertheless, there can be a sepcial unitary and $\rho_{\rm CTC}$ when the 
puification may be possible. 
For example, when  $f_m(k) = 1/d $ for all $k, m$ and $p_n = 1/d$ for all 
$n$ (the CTC density matrix is a random mixture), then 
Spec($\rho_{\rm CTC}$) $=$ Spec($\rho_{\rm CTC'}$) holds and 
purification of CTC density matrix is possible for arbitrary $\rho_{\rm CR}$.

 Therefore, in quantum world  
where CTCs exist we do have two kind of mixtures.
Even though practitioners of standard quantum theory do not think that there 
should be a difference between `proper' and `improper' mixtures, there are
some physicists and philosophers \cite{esp,hugh} who seem to believe that 
these two are distinct. 
They sometimes also discuss it in terms of objective versus subjective
probabilities. The purification not being possible for the CTC system
would imply that there could be objective
probabilities after all and presumably would say something about the
supposed reduction of classical probabilities to pure quantum evolution that
the many worlds people would like to believe.

Now, one might think that if the state of a CR system and a CTC system is 
already in a pure entangled state (say) 
$|\Phi \rangle_{\rm CR, CTC} = \sum_k \sqrt{p_k} |a_k \rangle_{\rm CR} 
|k \rangle_{\rm CTC}$, then 
by tracing out the CR system the state of the CTC subsystem will be in a 
mixed state. Now the question is can we always create such an entangled state 
between a CR and CTC system using any unitary? One can argue that this 
may not be the case.
Because if a CR and a CTC system is in a pure entangled state then at 
some point of time they must have interacted. Once they interact the CTC system
has to satisfy the consistency condition of Deutsch. One can check 
that this will lead to 
contradictions. Suppose we start with a CR system and a CTC system in pure 
product states, i.e., $\rho_{\rm CR} \otimes \rho_{\rm CTC} = 
|\psi\rangle_{\rm CR} \langle \psi | \otimes |\phi\rangle_{\rm CTC} 
\langle \phi |$ for some $|\psi\rangle \in {\cal H}_{\rm CR}$ and 
for some $|\phi\rangle \in {\cal H}_{\rm CTC}$. Then after the evolution 
the state of the CR and CTC 
system is  $|\Phi \rangle \langle \Phi |$ with 
$\rho_{\rm CTC} = \sum_k p_k | k \rangle \langle k |$. This shows that 
the consistency 
condition is not satisfied. Similarly, if we start with the initial 
$\rho_{\rm CR}$ as a pure state and the initial $\rho_{\rm CTC}$ as a mixed 
state, then the question is how can an overall mixed state (CR and CTC) 
evolve to a pure state  $|\Phi \rangle_{\rm CR, CTC}$ via unitary transformation.
In either case we reach a contradiction.  

So the question is how can one 
create entanglement between a CR system and a CTC system? It is possible 
to create entanglement between a CR system and a CTC system by first creating 
entanglement between two CR systems (in our world) and then swapping half of 
the CR subsystem with a CTC system whose density matrix is same as 
the reduced density matrix of the CR system. 
%Now note the state of CR system and CTC system both are in mixed states. 
To illustrate this clearly, 
let us consider two causality-respecting quantum systems in a pure 
entangled state 
$|\Psi \rangle_{\rm CR, CR'} = \sum_k \sqrt{p_k} |a_k \rangle_{\rm CR} 
|k \rangle_{\rm CR'}$. Let the initial state of $\rho_{\rm CTC} = 
\sum_k p_k |k \rangle_{\rm CTC} \langle k |$. Let half of the 
causality-respecting system ${\rm CR'}$
and the CTC system interact where the unitary operation is the swap operation.
After this interaction we will find that the state of the other half of the
CR system and the CTC system is actually in an pure entangled state 
$|\Psi \rangle_{\rm CR, CTC} = \sum_k \sqrt{p_k} |a_k \rangle_{\rm CR} 
|k \rangle_{\rm CTC}$ with the fixed point solution 
$\rho_{\rm CTC} = 
\sum_k p_k | k \rangle_{\rm CTC} \langle k |$. Now one may wonder, since 
the older version of the CTC system is same as the younger version of the 
CTC system and 
the younger version of the CTC state which was in a mixed state is actually now 
part of a pure entangled state how does it bypass the no-purification theorem?
Below we answer this question. 

{\bf Theorem II:} Let $\rho_{\rm CR'} = 
\sum_i \lambda_i |b_i \rangle_{\rm CR'} \langle b_i |$ and 
%be the density matrix of the causality-respecting system and 
$\rho_{\rm CTC} = 
\sum_k p_k |k \rangle_{\rm CTC} \langle k |$ are the density 
matrices of the causality-respecting system and CTC system, with their 
respective spectral decompositions.
Let them interact via an arbitrary unitary operator $U$ that acts on 
${\cal H}_{\rm CR'} \otimes {\cal H}_{\rm CTC}$. If $\rho_{\rm CTC}$ 
satisfies the consistency condition, then it cannot be purified for 
arbitrary $\rho_{\rm CR'}$ and $U$ such that  
$\rho_{\rm CTC} = 
{\rm Tr}_{\rm CTC'}(|\Phi\rangle_{\rm CTC,CTC'} \langle \Phi|)$ with 
$|\Phi \rangle_{\rm CTC,CTC'} = \sum_k \sqrt{p_k} |k \rangle_{\rm CTC} 
|d_k \rangle_{\rm CTC'}$.
%being indepedent of $\rho_{\rm CR'}$ and $U$.

{\bf Proof:}  Let the causality-respecting system that interacts with CTC 
system is part of a pure entangled states, i.e., $\rho_{\rm CR'} = 
{\rm Tr}_{\rm CR}(|\Psi\rangle_{\rm CR,CR'} \langle \Psi|)$ with 
$|\Psi \rangle_{\rm CR, CR'} = \sum_i \sqrt{\lambda_i} |a_i \rangle_{\rm CR} 
|b_i \rangle_{\rm CR'}$. Next, assume that we can purify the CTC system, i.e., 
$\rho_{\rm CTC} = 
{\rm Tr}_{\rm CTC'} (|\Phi\rangle_{\rm CTC,CTC'} \langle \Phi|)$ with 
$|\Phi \rangle_{\rm CTC, CTC'} = \sum_k \sqrt{p_k} |k \rangle_{\rm CTC} 
|d_k \rangle_{\rm CTC'}$. 
In the pure state picture, the Deutsch unitary evolution 
between the causality-respecting system and the chronology violating 
system can have the following transformation 

\begin{align}
%\begin{eqnarray}
 |\Psi \rangle_{\rm CR,CR'} |\Phi \rangle_{\rm CTC, CTC'}  \rightarrow 
(I \otimes U \otimes I) 
|\Psi \rangle_{\rm CR,CR'} |\Phi \rangle_{\rm CTC, CTC'}. \nonumber\\
%& =  \sum_{ik} \sqrt{ \lambda_i p_k} |a_i \rangle_{\rm CR} 
%U(|b_i \rangle_{\rm CR'} |k   \rangle_{\rm CTC} ) |d_k  \rangle_{\rm CTC'}.
%\end{eqnarray}
\end{align}
Thus, if the purification is possible then we have 

\begin{align}
& U[ \sum_i \lambda_i |b_i \rangle_{\rm CR'} \langle b_ i | 
\otimes\sum_k p_k |k \rangle_{\rm CTC} \langle k |  ]U^{\dagger} 
 \nonumber\\ 
& = \sum_{ik} \lambda_i p_k  U(|b_i \rangle_{\rm CR'} \langle b_i | \otimes 
|k \rangle_{\rm CTC} \langle k |))U^{\dagger}.
\end{align}
Let us define the action of the unitary operator on the basis of 
${\cal H}_{\rm CR'}$ and  ${\cal H}_{\rm CTC}$ as
\begin{align}
U(|b_i \rangle_{\rm CR'} |k \rangle_{\rm CTC} )= 
\sum_m \sqrt{g_m(i, k)} |B_m^k(i) \rangle_{\rm CR'} |C_m^i(k) \rangle_{\rm CTC}, 
\end{align}
where  $g_m(i,k)$'s are the Schmidt coefficients with $\sum_m g_m(i, k) =1$ 
for all $(i, k)$, $|B_m^k(i) \rangle_{\rm CR'}$ and 
 $|C_m^i(k) \rangle_{\rm CTC}$ are the Schmidt bases in their respective 
Hilbert spaces.  
%the output state of the combined system is given by 
%\begin{align}
%& (I \otimes U \otimes I) 
%|\Psi \rangle_{\rm CR,CR'} |\Phi \rangle_{\rm CTC, CTC'} \nonumber\\
%& =  
%\sum_{ikm} \sqrt{ \lambda_i p_k g_m(i, k)} |a_k \rangle_{\rm CR} 
%|B_m^k(i) \rangle_{\rm CR'} |C_m^i(k) \rangle_{\rm CTC}
%|d_k  \rangle_{\rm CTC'}.
%\end{eqnarray}
%\end{align}
%If we trace out CR, ${\rm CR'}$ and ${\rm CTC'}$ systems and use 
If we trace out ${\rm CR'}$ and use 
the kinematic consistency condition we have
\begin{align}
\sum_k p_k |k \rangle_{\rm CTC} \langle k| = 
\sum_{k} p_k \sum_{im} \lambda_i g_m(i,k) |B_m^k(i)\rangle_{\rm CTC} 
\langle B_m^k(i)|.
\end{align}
From (15), we have 
\begin{align}
p_n = \sum_{ikm} p_k \lambda_i g_m(i,k) |\langle n |B_m^k(i)\rangle |^2.
\end{align}
Thus, the egienvalues of $\rho_{\rm CTC}$ depends on $\rho_{\rm CR'}$ and 
$U$ whereas the egienvalues of $\rho_{\rm CTC'}$ are independent of the 
later. Therefore, if purification is possible for arbitrary
$\rho_{\rm CR'}$ and $U$ then the  
equal spectrum condition is violated. Hence, the proof. 

However, there are some special unitaries and CTC density matrix when 
purification can hold. For example, when the ${\rm CR'}$ system and 
the CTC system 
are subject to swap operation, i.e., $U= U_{\rm SWAP}$ 
and $\rho_{\rm CTC} = \rho_{\rm CR'}$.
In this process, entanglement is created between the 
CR and the CTC systems, and between the ${\rm CR'}$ and the ${\rm CTC'}$ 
systems. Also, general purification can hold when 
$g_m(i, k)$ and $p_n$ are equal (say $1/d$, i.e. for a random 
CTC density matrix) for all $i, k, m, n$.

In the Deutsch formalism, the mixed state $\rho_{\rm CTC}$ is
constrained by (2). Also 
this is in the form of a product state with the causality
respecting quantum system. This allows the state of the system to evolve
from a pure state to a mixed state, which is forbidden in the
standard quantum mechanics. One may be tempted to have an entangled pure
state picture as suggested by Deutsch where one takes into consideration 
of the many-worlds interpretation \cite{deu}. That is the CTC subsystem 
in our world can be regarded as being entangled with 
the CTC and CR subsystems in other world. However, our result rules out this
possibility (as the interacting CTC cannot be correlated with another world in
a universal manner).

To conclude, we have proved a no-purification theorem for quantum systems
in mixed states that traverse closed time like curves.
%and interact with causality respecting quantum systems
It is shown that given a
CTC system in a mixed state, if it has to evolve via Deutsch unitary gate 
and satisfy the kinematic consistency condition for all $\rho_{\rm CR}$ 
and all $U$ then it cannot be regarded as a subsystem of a universal pure
entangled state.  Thus, in general, there is no universal 
`Church of the larger Hilbert 
space' for CTC quantum system. However, as long as a CTC system does not 
interact with a CR system in a nontrivial way `improper' mixture may exist.
One may argue that for a fixed known $\rho_{\rm CR}$ and $U$ one can purify 
a CTC density matrix, but
the physical meaning of such a context dependent purification is not clear. 
In particular, if the other half of the CTC system is far away, then why 
should that depend on what is going on at a remote location.
Therefore, in quantum theory with closed time 
like curves, there can be two kinds of mixtures, namely, `proper' and 
`improper' mixtures. This also reveals the true nature of the density 
operators in quantum theory with CTCs. In the absence of CTCs, there 
is no way to distinguish them. If CTC can help in distinguishing `proper' from 
`improper' mixture then it may lead to signalling. How exactly this can 
happen will be reported in future. In essence 
our result brings out a very fundamental and important difference 
between the density matrix of the CR system and the CTC system. 
It is likely that the nature of entanglement between causality-respecting 
system and CTC system may be different. We hope that our result will add new 
insights to quantum information theory 
in the presence of closed time like curves.
Much more remains to be explored in this direction.

\vskip 2 cm

{\it Acknowledgment:} We thank S. L. Braunstein, G. Chiribella 
and G. Kar for useful comments. We also thank G. Smith for his 
critical comments.


\begin{thebibliography}{99}
\bibitem{akp}
%We can also understand why purification fails in quantum theory with CTCs. 
In the Deutsch model because of the fixed point condition 
the density matrix $\rho_{\rm CTC}$ is a nonlinear function of the input 
state $\rho_{\rm CR}$. This makes the whole evolution nonlinear. 
In a nonlinear theory the purification 
equation (7) may not hold in general. That it is violated can be seen 
from the fact that the evolution of a mixture is not equal to the mixture 
of the evolutions, in general, i.e.,  
%That is to say that in nonlinear theory the following equation 
%cannot hold for the evolution of a general density matrix
%\begin{align}
$ U (\sum_k \lambda_k \phi_k \otimes \rho_k)U^{\dagger} \not = 
\sum_k p_k U( \phi_k \otimes \rho_k )U^{\dagger} $.
%\end{align}
%If we take $\rho_k = |\psi \rangle \langle \psi| \otimes 
%|k \rangle \langle k $, then  the violation of (7) follows from (8).
%The action of $U$ on superposition of vectors does not define the action 
%on mixtures. 
We conjecture that the no purification theorem may be true even in other 
nonlinear quantum theories.
\end{thebibliography}
\end{document}